\documentclass{article}

\usepackage[final]{mlsafety_neurips_2022}

\usepackage[utf8]{inputenc} 
\usepackage[T1]{fontenc}    
\usepackage{hyperref}       
\usepackage{url}            
\usepackage{booktabs}       
\usepackage{amsfonts}       
\usepackage{nicefrac}       
\usepackage{microtype}      
\usepackage{xcolor}         
\usepackage{graphicx}

\usepackage{array}
\usepackage{amssymb}
\usepackage{enumitem}
\usepackage{xcolor}
\usepackage{soul}

\title{A Multi-Level Framework for the AI Alignment Problem}

\author{
    Betty Li Hou\textsuperscript{1, 2}, Brian Patrick Green\textsuperscript{1} \\
    \textsuperscript{1} Markkula Center for Applied Ethics \\
    \textsuperscript{2} New York University \\
    \texttt{blh9134@nyu.edu, bpgreen@scu.edu}
}

\begin{document}
    \maketitle

\begin{abstract}
  AI alignment considers how we can encode AI systems in a way that is compatible with human values. The normative side of this problem asks what moral values or principles, if any, we should encode in AI. To this end, we present a framework to consider the question at four levels: Individual, Organizational, National, and Global. We aim to illustrate how AI alignment is made up of value alignment problems at each of these levels, where values at each level affect the others and effects can flow in either direction. We outline key questions and considerations of each level and demonstrate an application of this framework to the topic of AI content moderation.
\end{abstract}

\section{Introduction}

AI is used on a global scale in a multitude of ways, from social media algorithms and cybersecurity to smart home devices and increasingly-more-autonomous vehicles. This poses risks of both direct and indirect negative effects on our political, economic, and social structures. With this new realm of technology, we must thoroughly understand and work to address the risks in order to navigate the space and use the technology wisely. This is the field of AI ethics, and here, specifically, AI safety.

\subsection{AI Alignment}

The AI alignment problem considers how we can encode AI systems in a way that is compatible with human moral values. The problem becomes complex when there are multiple values that we want to prioritize in a system. For example, we might want both speed and accuracy out of a system performing a morally relevant task, such as online content moderation. If these values are conflicting to any extent, then it is impossible to maximize for both. AI alignment becomes even more important when the systems operate at a scale where humans cannot feasibly evaluate every decision made to check whether it was performed in a responsible and ethical manner.

The alignment problem has two parts [1]. The first is the technical aspect which focuses on how to formally encode values and principles into AI so that it does what it ought to do in a reliable manner. Cases of unintended negative side effects and reward hacking can result if this is not done properly [2]. The second part of the alignment problem is normative, which asks what moral values or principles, if any, we should encode in AI. To this end, we present a framework to consider the question at four levels.\footnote{For another take on the problem, see the “multiscale alignment” section of Max Tegmark’s interview with the 80,000 Hours Podcast [3]. Tegmark’s framework does not yet seem to be published, so we cannot know in exactly what ways his and our frameworks are similar or different.} While other notable recent papers have focused more on the content of the \newpage problem solution (e.g. Hendrycks et al., 2021, [4] and Hendrycks et al., 2022, [5]), this framework focuses on the social context in which those content-focused solutions must exist, and how that context has multiple layers across which solutions should be integrated and coherent.

\section{Breaking Down the Alignment Problem}
AI alignment is made up of value alignment problems at multiple different levels, not just in the technology itself, how it is built, and the design methods. In order for AI to truly be aligned with human moral values, all levels must be aligned with each other as well. The following is an approach to AI alignment in which the values at each level affect the others, with effects flowing both downwards and upwards. At each level, there are key questions that need to be answered.

\begin{figure}[h]
    \centering
    \includegraphics[width=0.7\linewidth]{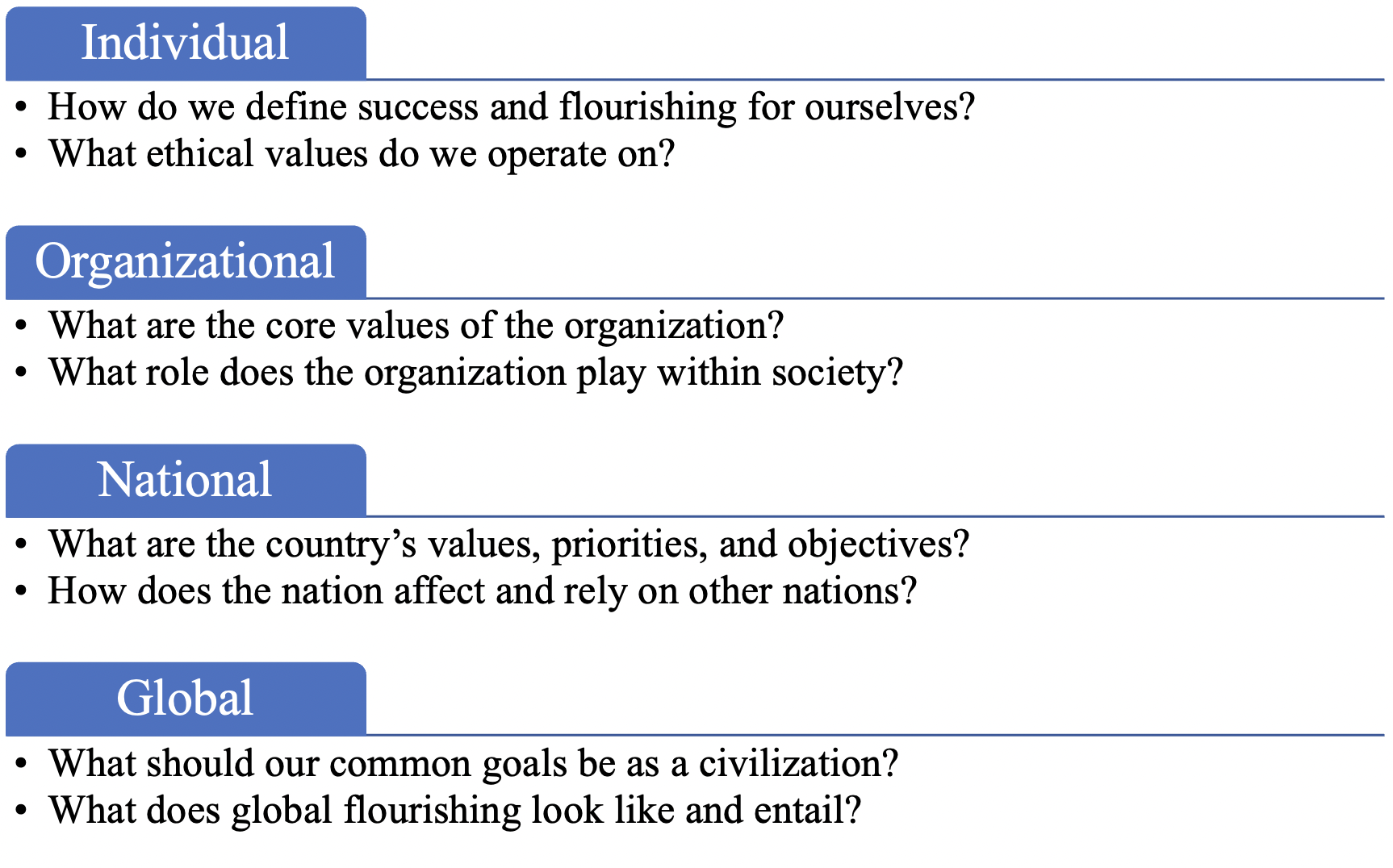}
    \caption{Four Levels of Alignment}
    \label{fig:my_label}
\end{figure}

\paragraph{Individual \& Familial}
On the individual level, the framework invites individuals and families to ask questions about values and flourishing. In our everyday actions, we are shaping our own definitions of individual flourishing—what makes life fulfilling and brings contentment. We must consider what role models and lifestyles we seek to emulate, how we define success for ourselves, what sacrifices we are willing to make, and what ethical values we prioritize.

\paragraph{Organizational}
The organizational level refers to corporations, state and local governments, universities, churches, social movements, and various other groups in civil society. When considering alignment at this level, we must determine what values the organization operates on, what values are instilled in its products and services, and what role the organization plays within society. For institutions, important considerations are what constitutes success, what metrics are used to evaluate success, and how they are involved in the broader movements for AI alignment.

\paragraph{National}
The next level is the national level. Each nation has either implicitly or explicitly defined values that determine the country’s goals and objectives pertaining to AI. A country aiming to assert itself as a global power may invest resources into building a domestic AI industry, as well as regulate the usage of AI to moderate and nudge users’ behaviors towards particular views. On the other hand, a country aiming to promote freedom may follow a decentralized approach to AI production, giving firms freedom and privacy while allowing for competition amongst firms. Alternatively, countries may try to build an AI initiative in a way that not only ensures that they are aligned with moral values, but also encourages or requires other countries to do so.

\paragraph{Global}
Globally, humankind must think about the kind of future we want to have. The recently articulated United Nations Sustainable Development Goals (SDGs) offer a good starting point, but these goals are merely the preconditions necessary for survival and flourishing, so they are not enough [6]. A further step is needed to determine our common goals as a civilization, and more philosophically, the purpose of human existence, and how AI will fit into it. Is it to survive, raise children, live in society, seek the truth, etc.? Related to this are the end goals of economic and political structures, as well as what powerful nations and corporations need to give up in order to attend to the needs of the poor and the earth.

\subsection{Putting the Levels Together}
All of these levels interact with each other. Because AI typically originates from the organizational level, often in profit driven corporations, the primary motivation is often simply to make money. However, when put in the context of these other levels, further goals should become visible: 1) AI development should be aligned to individual and familial needs, 2) AI development should align with national interests, and 3) AI development should contribute to human survival and flourishing on the global level. 

But other layers in the framework also interact with each other, through inputs and outputs. For example, looking at the same organizational layer from the inbound perspective, individuals can choose whether or not to buy certain kinds of technologies, nations can pass laws and regulations to control what technology companies can do, and at the global level, international pressure (for example from the UN through ideas such as the Sustainable Development Goals) can also influence technology company behavior. Of note, these levels can have intermediate levels too, such as the European Union, which is above national but below global, and which has, through the General Data Protection Regulation (GDPR) [7], had a major influence on the internet, data, and through those, AI.

Examining the individual level, we have already seen how it influences and is influenced by the organizational level. The individual level can influence the national through elections, and the global through organizations such as the UN, although these influences are quite underdeveloped. Similarly, the global can influence individuals through international treaties, while nations obviously exert significant control over their citizens through laws and other behavioral expectations.

Lastly, the national and global levels interact. Nations influence the global state of the Earth, for example through war and other national policies with global effects (such as energy policies which can drive or mitigate climate change.) The global level can exert power back, whether through the UN or other international expectations of national behavior. 

To get a more practical view of the framework, we look at the problem of social media content moderation.

\section{Content Moderation as an Example}
A global debate has emerged on the risks posed by certain types of content on the internet. User generated content is not subject to the same editorial controls as traditional media, which enables users to post content that could harm others, particularly children or vulnerable people. This includes but is not limited to content promoting terrorism, child abuse material, hate speech, sexual content, and violent or extremist content. Yet at the same time, attempts to restrict this content can seem to some like it violates user freedom of expression and freedom to hear certain kinds of expression. Organizations and governments have grappled with the feasibility and ethics of mitigating these potential harms through content moderation, while at the same time trying not to lose users who feel that their freedoms are being curtailed.

AI-assisted content moderation brings a level of speed and scale unmatched by manual moderation. A transparency report from Google shows that over 90\% of videos removed on YouTube between April and June 2022 were reviewed as a result of automatic flagging [8]. However, these approaches have implications for people’s future uses and attitudes towards online content sharing, so it is important that the AI employed in these processes aligns with human values at multiple levels.

\subsection{Using the Framework}

\paragraph{Organizational $\Rightarrow$ Individual}
The first issue comes from the organizational level, where there is a major misalignment between businesses and individuals. Businesses that employ content moderation are incentivized to maximize shareholder value, which leads to prioritizing profit over social good. For example, companies often base their algorithm on “engagement”—the more likes, comments and shares a topic or post receives, the more it will appear on people’s newsfeeds. Per profile as well, companies then keep track of the user’s behavior and habits based on engagement to feed them what they want to see. This way, users will spend more time on the site and generate more ad revenue for the business to boost shareholder value. This however leads to echo chambers and polarization, as users are not exposed to opinions that differ from theirs, ultimately affecting not only individuals and families, but also entire nations, and even global discourse. The misalignment between organization and individuals has already proven to be dangerous with cases like Myanmar’s attack on minorities illustrating the potential consequences [9].

\paragraph{National $\Rightarrow$ Organizational $\Rightarrow$ Individual}
National regulations shape how organizations moderate content, as organizations must build AI within the bounds of these regulations. A country’s content moderation legislation is typically an expression of the cultural values of the majority of its citizens, which is often similar to the cultural values of its leadership, though not always. While these regulations are made by individual lawmakers and may express the values of many individual citizens, these regulations also will affect both organizations and other individuals. For example, a common good perspective might lean towards high content moderation for the sake of minimizing social harm, but at the expense of individual freedom of expression. 

The question then arises regarding the alignment of cultural values with AI content moderation. We may be able to recognize where there are mis-alignments between national and organizational values, which in turn affects individuals. For example, in the US, where individual freedoms is a priority, there is very little content moderation regulation and it requires companies to only moderate things such as illegally sharing copyrighted content and criminal activity such as sharing child sexual abuse materials. Therefore, while companies may comply with every relevant government regulation, there have nevertheless been harmful effects on society, showing how the US government content moderation legislation is not aligned with societal needs. Cases like Myanmar also suggest that this American legislation may not be aligned with global needs, as other countries are subject to these same problems and are facing the repercussions of it. 

Based on the above, it might seem that the first goal for AI alignment would be to align the national and organizational levels (assuming that the organization is also aligned with individual well-being). However, this is not enough—we must also consider whether these national values are aligned on the global level, that is, whether they support global human flourishing. Content which promotes division, subversion of governments, extremism, conspiracy theories, and other socially destructive and anti-social behaviors can result in not only local or national problems, but, if these groups become broadly empowered, global problems. And yet governing content moderation at the international level is simply not feasible in our pluralistic world. However, this framework at least helps to diagnose the problem, even if the problem cannot yet be solved.

\paragraph{Individual $\Rightarrow$ Organizational $\Rightarrow$ National $\Rightarrow$ Global}
The effects flow in both directions. Organizations doing content moderation sometimes respond most to individual user feedback, a powerful enough organization can have a hand in swaying national interests, and a nation or group of nations can potentially change the international order, for example, as with the EU's GDPR.

All in all, content moderation is a prime example of how value alignment is at work right now in society. It may not be feasible to align all four levels quickly or easily, but with this framework we can identify some causes of the complex mis-alignments that we are seeing now—and will see more of in the future—and consider some possible beginnings of solutions.

\section{Conclusion}
If we are to make good progress on the normative side of AI alignment, we must consider all levels: Individual, Organizational, National, and Global, and understand how each works together, rather than only aligning one or a few of the parts. Here we have presented a framework for considering these issues. In Appendix \hyperref[firstappendix]{A}, we have provided an analysis of how this work relates to AI x-risk.

The versatility of the framework means that it can be applied to many other topics, including but not limited to autonomous vehicles, AI-assisted clinical decision support systems, surveillance, and criminal justice tools. In these hotly contested spaces with no clear answers, by analysing these problems at four levels, we are able to see the many interacting parts at play, in order to create more ethical and aligned AI.

\newpage
\section*{References}
{
\small

[1] Iason Gabriel. "Artificial Intelligence, Values, and Alignment". In: {\it Minds \& Machines}, 30:411–437, 2020.

[2] Dario Amodei, Chris Olah, Jacob Steinhardt, Paul Christiano, John Schulman, and Dan Mané. "Concrete Problems in AI Safety". In: {\it arXiv preprint arXiv:1606.06565}, 2016. 

[3] Robert Wiblin and Keiran Harris. “Max Tegmark on how a ‘put-up-or-shut-up’ resolution led him to work on AI and algorithmic news selection”. The 80,000 Hours Podcast, July 1st, 2022, minutes 1:13:13-1:51:01. URL: \texttt{https://80000hours.org/podcast/episodes/max-tegmark-ai-and-algorithmic-news-selection/}.

[4] Dan Hendrycks, Collin Burns, Steven Basart, Andrew Critch, Jerry Li, Dawn Song, and Jacob Steinhardt. "Aligning AI with Shared Human Values". In: {\it arXiv preprint arXiv:2008.02275}, 2021. 

[5] Dan Hendrycks, Nicholas Carlini, John Schulman, and Jacob Steinhardt. "Unsolved Problems in ML Safety". In: {\it arXiv preprint arXiv:2109.13916v5}, 2022.

[6] "THE 17 GOALS - Sustainable Development Goals".  United Nations. URL: \texttt{https://sdgs.un.org/ \ goals}.

[7] "General Data Protection Regulation". European Unions. URL: \texttt{https://gdpr-info.eu}.

[8] “YouTube Community Guidelines enforcement”. Google. URL: \texttt{https://transparencyreport.google. \ com/youtube-policy/removals}.

[9] Nathaniel Persily and Joshua A. Tucker. "Social Media and Democracy: The State of the Field, Prospects for Reform". Cambridge University Press, 2020.

}

\appendix
\section{X-Risk Analysis} \label{firstappendix}

AI X-Risk Analysis template from: Dan Hendrycks and Mantas Mazeika. "X-Risk Analysis for AI Research". In: \textit{arXiv preprint arXiv:2206.05862v7}, 2022.

\subsection{Long-Term Impact on Advanced AI Systems}
In this section, we analyze how this work shapes the process that will lead to advanced AI systems and how it steers the process in a safer direction.
\begin{enumerate}[leftmargin=*]
\item \textbf{Overview.} How is this work intended to reduce existential risks from advanced AI systems? \\ 
\textbf{Answer:} Advanced AI systems will have to exist within the context of the world composed of different nations, organizations, and individuals. In order to avoid global catastrophe, we must look at all the levels. By aligning these and considering the ways in which they interact, we can be more confident that advanced AI systems can be built to be compatible with the world, or at least be more aware of the ways in which they can potentially cause conflict. In this way, developers can have a better idea of how they aim to have the system operate in the world. But without doing the former, there is no way for future advanced AI, including AGI, to be aligned in any sense on a global scale, which opens the door for existential risk. These misalignments are already visible when it comes to AI content moderation which has yielded innumerable ill social effects in the hands of malicious, deluded, or merely ignorant actors.

\item \textbf{Direct Effects.} If this work directly reduces existential risks, what are the main hazards, vulnerabilities, or failure modes that it directly affects? \\ 
\textbf{Answer: }
The work directly addresses the problem of existential risk by clarifying the preconditions for any real solutions to the problems posed by X-risks. Proposed solutions which do not consider how the individual, organization, national, and global levels integrate will be flawed, and therefore not real solutions. This work highlights how we should go about directly mitigating risks: first, by understanding how the levels align or misalign, and then considering how AI solutions to risk are likely to exist within that multilevel structure. A failure mode of this is that, as knowledge, it could be used for creating misalignment as well, allowing for optimizing multilevel misalignment rather than alignment. Certain nations and groups have already exploited AI moderated systems in this way, so this is hardly a new discovery on the negative side—it is the positive side which seems to have remained implicit until now, and so, by making it explicit, hopefully we can begin to recapture the escaped genie of misaligned AI and reverse its effects.

\item \textbf{Diffuse Effects.} If this work reduces existential risks indirectly or diffusely, what are the main contributing factors that it affects? \\ 
\textbf{Answer: } 
The main effect of this paper is a diffuse reduction in X-risk, by enabling those who generate other AI alignment X-risk solutions to better understand the multilevel nature of society in which solutions must exist, and then how to integrate their solutions into that multilevel society. If the ideas in this paper are ignored then AI alignment X-risk solutions are likely to be less effective and less comprehensive, thus not solving the AI alignment problem as well as they could have if the ideas in this paper had been more fully integrated with their work. It is not enough to merely create alignment at one level, for example, that of government, since various governments of the world will remain misaligned. Currently we see this between autocracies vs. democracies and free nations vs. oppressive ones. AI alignment X-risk solutions which merely enable nations to align their populaces with the government will only empower oppressive autocracies and make global misalignment worse, not better. This is true for the other levels as well. Alignment has to be consistent from top to bottom or the problem merely squeezes out into the other levels, like a water balloon about to burst.

\item \textbf{What’s at Stake?} What is a future scenario in which this research direction could prevent the sudden, large-scale loss of life? If not applicable, what is a future scenario in which this research direction be highly beneficial? \\
\textbf{Answer: }
As discussed previously, this framework could help to prevent situations like what happened in Myanmar as a result of AI-assisted social media content moderation. If the businesses creating these systems are aligned with individual, national, and global needs, we can begin to fix the problems of polarization, echo chambers, and remove the possibility for AI to be exploited the way it was in Myanmar to target minorities. Without this work, a theoretical AGI could be built in line with one country's values but in direct conflict with another country's values and cause significant harm, or have to be restricted to a smaller domain, say just one country. However, if the organizations in a country are aligned to one definition of success, and countries are united on a definition of global flourishing, then it is possible to create an AGI that could provide much greater social benefit. 

\item \textbf{Result Fragility.} Do the findings rest on strong theoretical assumptions; are they not demonstrated using leading-edge tasks or models; or are the findings highly sensitive to hyperparameters? \hfill $\boxtimes$

\item \textbf{Problem Difficulty.} Is it implausible that any practical system could ever markedly outperform humans at this task? \hfill $\square$

\item \textbf{Human Unreliability.} Does this approach strongly depend on handcrafted features, expert supervision, or human reliability? \hfill $\square$

\item \textbf{Competitive Pressures.} Does work towards this approach strongly trade off against raw intelligence, other general capabilities, or economic utility? \hfill $\square$

\end{enumerate} \subsection{Safety-Capabilities Balance}
In this section, we analyze how this work relates to general capabilities and how it affects the balance between safety and hazards from general capabilities.

\begin{enumerate}[resume,leftmargin=*]
\item \textbf{Overview.} How does this improve safety more than it improves general capabilities? \\ 
\textbf{Answer: }
The paper frames how any efforts to improve general capabilities should first consider safety in the broader societal context and how to do so. Because this is a framework within which thinking about general capabilities exists, on the downside, those capabilities could be directed towards bad ends in a more comprehensive way—however, it seems that the nations who act as spoilers in international affairs already understand this, and so it is the other nations of the world, and organizations and individuals, who wish to improve the state of the world that need to work harder to learn this lesson and act to align AI in a more comprehensive and integrated way. Therefore this promotes safety more than hindering it.

\item \textbf{Red Teaming.} What is a way in which this hastens general capabilities or the onset of x-risks? \\ 
\textbf{Answer:}
This hastens general abilities in AI only in the same way as any comprehensive model of society might enable better thinking about the integration of technology into society. However, it could perhaps be used by bad actors to either attempt to sow conflicting alignments across various AIs, or subtly direct AI towards one very bad misalignment. If an AGI were to be built in these ways, then safety would be harmed—but this appears to be almost the default mode for current AI work (often confused, not aligned between levels, or intentionally abused by malicious actors), and so by being more explicit about the true form of the problem, it would seem that the likely direction of progress among those with good intent would be towards improvement and not towards degradation.

\item \textbf{General Tasks.} Does this work advance progress on tasks that have been previously considered the subject of usual capabilities research? \hfill $\square$

\item \textbf{General Goals.} Does this improve or facilitate research towards general prediction, classification, state estimation, efficiency, scalability, generation, data compression, executing clear instructions, helpfulness, informativeness, reasoning, planning, researching, optimization, (self-)supervised learning, sequential decision making, recursive self-improvement, open-ended goals, models accessing the Internet, or similar capabilities? \hfill $\boxtimes$

\item \textbf{Correlation With General Aptitude.} Is the analyzed capability known to be highly predicted by general cognitive ability or educational attainment? \hfill $\square$

\item \textbf{Safety via Capabilities.} Does this advance safety along with, or as a consequence of, advancing other capabilities or the study of AI? \hfill $\boxtimes$ 

\end{enumerate}

\subsection{Elaborations and Other Considerations}

\begin{enumerate}[resume,leftmargin=*]
\item \textbf{Other.} What clarifications or uncertainties about this work and x-risk are worth mentioning? \\ 
\textbf{Answer:} 
This paper is a description of and framework for the context in which AI-associated risks appear. As a description and framework, it allows the problem to be understood in a new way, and one which is hopefully enlightening to those seeking to solve the AI alignment problem. Because the alignment problem runs on a continuous spectrum from contemporary problems like content moderation, all the way to AGI and X-risks, this framework should provide helpful insights to those seeking to create comprehensively safe AI. And for those who would seek to do the opposite, they already know how to sow discord and evil. This framework is a tool for those who seek to do good, better. The dual use is essentially useless for those motivated by malice, as it is already well-known by them at a tactical, operational, and strategic level. It is those on the side seeking the comprehensive good that should find this to be of most help, tactically, operationally, and strategically, making AI safer for everyone.
\end{enumerate}

\end{document}